\documentclass[12pt]{iopart}
\usepackage{iopams,epsfig}  
\def\sqr#1#2{{\vcenter{\hrule height.#2pt\hbox{\vrule width.#2pt
height#1pt \kern#1pt \vrule width.#2pt}\hrule height.#2pt}}}

\def\hook{\hbox{\vrule height0pt width4pt depth0.3pt
\vrule height7pt width0.3pt depth0.3pt \vrule height0pt width2pt
depth0pt} }

\def\brr{\begin{eqnarray}}
\def\err{\end{eqnarray}}
\def\brn{\begin{eqnarray*}}
\def\ern{\end{eqnarray*}}
\def\beq{\begin{equation}}
\def\eeq{\end{equation}}
\def\vt{\vartheta}

\def\a{\alpha}
\def\b{\beta}
\def\m{\mu}
\def\n{\nu}

\def\d{\delta}

\def\r{\rho}
\def\L{{\mathcal{L}}}
\def\T{{\mathcal{T}}}

\def\C{{\mathcal{C}}}
\def\F{{\mathcal{F}}}

\eqnobysec

\begin{document}
\title{Energy-momentum current for coframe gravity }
\author{Yakov Itin}
\address{Institute of Mathematics,  
Hebrew University of Jerusalem\\
 Givat Ram, Jerusalem 91904, Israel, 
itin@math.huji.ac.il}
\begin{abstract}
The obstruction for the existence of an energy momentum tensor for  the 
gravitational field is  connected with  differential-geometric features 
of the Riemannian manifold.   
It has not to be valid  for alternative geometrical structures.\\
A teleparallel manifold is defined as a parallelizable
differentiable $4D$-manifold  endowed with a class of smooth coframe fields 
related by global Lorentz, i.e., $SO(1,3)$ transformations.
In this article a general free parametric class of teleparallel models 
is considered.
It includes  a 1-parameter subclass of viable models with the Schwarzschild
coframe solution.\\
A new form of the coframe field equation is derived  from the general 
teleparallel Lagrangian
by introducing the notion of a 3-parameter conjugate field strength $\F^a$.
The field equation turns out to have a form completely similar
to the Maxwell field equation $d*\F^a=\T^a$. 
By applying the Noether procedure, the source 3-form  $\T^a$ is shown to be 
connected with the diffeomorphism invariance of the Lagrangian. 
Thus the source $\T^a$ of the coframe field 
is interpreted as the total conserved energy-momentum current.  
The energy-momentum tensor  for coframe is defined. 
The total energy-momentum current of a system of a coframe and a 
material fields is conserved. 
Thus a redistribution of the energy-momentum current between a material 
and a coframe (gravity) fields is possible in principle, unlike as in 
the standard GR.\\
For special values of parameters, when the GR is reinstated, the energy-momentum 
tensor gives up the invariant sense, i.e., becomes a pseudo-tensor. 
Thus even a small-parametric change of GR turns it into a well 
defined Lagrangian theory. 

\end{abstract}
\pacno{ 02.40.Hw, 04.20.Cv }

\section{Introduction}          
The concept of an energy-momentum tensor  
for the gravitational field is, 
undoubtedly, the most puzzling issue in general relativity (GR). 
Such a tensor of non-geometric material fields acting 
in a  fixed  geometrical background is well defined.  
This quantity  (denote it by ${T^\mu}_\nu$) obeys the following properties. 
It is  
\begin{itemize}  
\item[(i)]
{\it local} - i.e., constructed only from the fields taken at an arbitrary point 
on a manifold and from the derivatives  of these fields taken 
at the  same point,
\item[(ii)]
{\it diffeomorphic covariant} - i.e.,   transforms as a tensor under   
diffeomorphisms of the manifold,
\item[(iii)] 
{\it inner invariant} - i.e.,  does not change under inner symmetry 
transformations of fields, which preserve  the Lagrangian, 
\item[(iv)] 
{\it conserved} - i.e., satisfies the covariant divergence equation 
${T^\mu}_{\nu;\mu}=0$, 
\item[(v)] 
{\it ``the first integral of the field equation''} - 
it is derivable from the field equations by integration. 
The order of field derivatives in ${T^\mu}_\nu$ is  of one less than 
the order of the field equation.
\end{itemize}
It is well known that in GR an energy-momentum tensor of the metric 
(gravitational) field itself,  satisfying the conditions  listed above, fails to  exist. 
This fact is usually related to  the equivalence principle, which 
implies that the gravitational field can not be detected at a point 
as a covariant object. 
This conclusion can also be  viewed  as a purely differential-geometric fact. 
Indeed, the components of the metric tensor  are managed by a system 
of second order partial differential equations.
Thus the energy-momentum
quantity has to be a local tensor constructed from the metric components and 
their first order derivatives. 
The corresponding theorem of  (pseudo) Riemannian geometry 
states that  every expression  of such a type is trivial. 
Thus, the objection for the existence of a gravitational 
 energy-momentum tensor is 
directly related to the geometric properties of the (pseudo) 
Riemannian manifold. 
It is natural to expect that  this objection can be lifted 
in an alternative model of gravity, even connected with the geometry 
of the manifold.\\
In resent time {\it teleparallel  structures}  on spacetime 
have evoked  a considerable  interest  for various reasons. 
They was considered as an essential part of generalized non-Riemannian theories 
such as the Poincar{\'e} gauge theory 
(see  \cite{H-N} - \cite{Blag2} and the references therein) or metric - affine gravity 
\cite{hehl95}
as well as a possible physical relevant geometry by itself - teleparallel description of gravity (see Refs. \cite{{Nitsch}} to \cite{Per}) . 
Another important subject are the various applications of the frame technique 
in physical theories based on classical (pseudo) Riemannian geometry. 
For instant in \cite{Mielke} teleparallel approach used for positive-gravitational-energy proof.  
In  \cite{T-N} the relation between spinor Lagrangian and teleparallel theory is established.\\
The most important property of the teleparallel theory is the existence of a family 
of viable gravitational models. \\
In the present paper we study the general free-parametric model on differential manifold, 
endowed with a metric constructed from the coframe. 
We start with a brief survey of the coframe teleparallel approach to  gravity. \\
Our main results are presented in the third section. 
We consider a pure coframe field with the most general 
odd quadratic coframe Lagrangian, which involves  
3 dimensionless parameters $\rho_1,\rho_2,\rho_3$. 
The field equation is derived in a form almost literally similar 
to the Maxwell-Yang-Mills  field equation. 
The source term of the equation is a conserved vector valued 3-form.  \\
By applying the Noether procedure, this 3-form 
is associated with the diffeomorphism invariance of the 
Lagrangian is derived. 
Hence, it is interpreted as the energy-momentum current of the coframe field. 
The notion of the Noether 
current and the Noether charge for the coframe field are introduced. \\
The energy-momentum tensor  is defined as a map of the module of
current 3-forms into the module of vector fields. 
Thus, the  energy-momentum tensor for the coframe field is 
defined in a diffeomorphism invariant and a translational covariant way. \\
For a system of a coframe and a material field it is shown that the total 
energy-momentum current serves as a source of the coframe field. 
This total current is conserved. 
Consequently, a redistribution  of energy between material 
and gravitational (coframe) fields is possible in principle.  \\
We briefly discuss the special case of the teleparallel equivalent of GR. 
This models turns to be an alternative formulation of GR, not an alternative model. 
The energy-momentum current in this case loses the invariant sense, with accordance with 
numerous investigations in standard GR.\\ 
Consequently, all the  viable teleparallel models with $\rho_1=0$, except of the  GR, 
have a well-defined energy-momentum tensor.    
\section{Teleparallel gravity}            
Let us give a brief account of gravity on teleparallel manifolds.  
Consider a coframe field $\{\vt^a, \ a=0,1,2,3\}$ defined 
on a $4D$  differential manifold $M$.
The 1-forms $\vt^a$ are declared to be pseudo-orthonormal.
This determines completely  a metric on the manifold $M$ by 
\begin{equation}\label{3.1} 
g=\eta_{ab}\vt^a\otimes\vt^b.
\end{equation}
So, the coframe field $\vt^a$ is considered as
a basic dynamical variable  while the metric $g$ is  treated as only
a secondary structure.\\ 
Such simple coframe structure is not complying with the relativistic paradigm 
because the coframe 1-forms $\vt^a$ produce peculiar directions at every point on $M$. 
In order to  have an isotropic structure the coframe variable have to  
be defined only up to
{\it global pseudo-rotations}, i.e. $SO(1,3)$ transformations. 
Consequently, the truly dynamical variable is  the equivalence
class of coframes   $[\vt^a]$, while 
the global pseudo-rotations produce  the  equivalence relation on this class. 
Hence, in addition to the invariance under the diffeomorphic transformations
of the manifold $M$, the basic geometric structure
has to be  global $SO(1,3)$ invariant. \\
The well known property of the teleparallel geometry 
is the possibility to define the parallelism of two vectors
at different points by 
comparing the components of the vectors in local frames. 
Namely, two vectors (1-forms) are parallel if the corresponding
components referred to a local frame (coframe) are proportional.
This  {\it absolute  parallelism} structure produces a global
path independent parallel transport.
In the affine-connections formalism such a transport is described 
by existence of a special
teleparallel connections of vanishing curvature \cite {hehl95}.
However, the Riemannian curvature of the manifold,  
which is constructed from the metric (\ref{3.1}),  is non-zero, in general. \\
Gravity is described by the teleparallel geometry
in a way similar to Einstein theory, i.e.,  by differential-geometric 
invariants of the structure. 
Looking for such invariants,  an important distinction between 
the metric and the teleparallel structures emerges.\\
{\it The metric structure} admits diffeomorphic invariants only of the second 
order or greater. 
The metric invariants of the first order are trivial. 
The unique invariant of the second order is the scalar curvature. 
This expression is well known to play the 
role of an  integrand in the Einstein-Hilbert action.\\
{\it The teleparallel structure} admits diffeomorphic and $SO(1,3)$ global 
invariants even of the first order. 
A simple example is the expression $e_a\hook d\vt^a$. 
The diffeomorphic invariant and global 
covariant operators, which can contribute to a general field equation, 
constitute a rich class \cite{i-k}. \\
Restrict the consideration to  
odd, quadratic (in the first order derivatives of the coframe field 
$\vt^a$),  diffeomorphic, and global $SO(1,3)$ invariant Lagrangians.  
A general Lagrangian of such type is represented by a linear combination 
of three Weitzenb\"{o}ck  quadratic  teleparallel invariants.  
The symmetric form of this Lagrangian  is \cite{Hehl98} ($\ell=$ Planck
length)
\begin{equation}\label{3.2}         
L^{cof}=
\frac 1{2\ell^2}\sum_{i=1}^3 \rho_{i} \; {}^{(i)}L, 
\end{equation}
with 
\brr\label{3.3}   
{}^{(1)}L &=&d\vt^a \wedge *d\vt_a,\\
\label{3.4}   
{}^{(2)}L &=&
\Big(d\vt_a \wedge \vt^a \Big) \wedge*\Big(d\vt_b\wedge\vt^b\Big), \\
\label{3.5}     
{}^{(3)}L &=& 
(d\vt_a \wedge\vt^b ) \wedge *\Big(d\vt_b \wedge \vt^a \Big).
\err
The 1-forms $\vt^a$ are assumed to carry the dimension of length, while 
the coefficients $\r_i$ are dimensionless. 
Hence the total Lagrangian $L^{cof}$ is  dimensionless. 
In order to simplify the formulas below we will use the Lagrangian 
$L=\ell^2 L^{cof}$ of the dimension: length square. 
In other worlds  the geometrized units system $G=c=\hbar=1$ is applied. 
For comparison with the ordinary  units see \cite{Wald-book}. \\
Every term of the Lagrangian (\ref{3.2}) is independent of  
a specific choice of a coordinate system and invariant under a global 
(rigid) $SO(1,3)$ transformation of the coframe. 
Thus, different choices of the free parameters $\r_i$ yield 
different translational and diffeomorphic invariant classical field models. 
Some of them are known to  be applicable for description  of  gravity.\\
The field equation is derived from the Lagrangian (\ref{3.2}) 
in the form \cite{Kop},\cite{Hehl98}
\brr\label{3.6}    
&&\rho_1\Big(
2d*d\vt_a+
e_a\hook(d\vt^b\wedge*d\vt_b)-
2(e_a\hook d\vt^b)\wedge*d\vt_b\Big)+\nonumber\\
&&\rho_2\Big(
-2\vt_a \wedge d *(d\vt^b\wedge \vt_b)+
2d\vt_a \wedge * ( d\vt^b \wedge \vt_b)+\nonumber\\
&&
e_a\hook\Big(d\vt^c\wedge\vt_c\wedge*(d\vt^b\wedge\vt_b)\Big)-
2(e_a\hook d\vt^b)\wedge\vt_b\wedge*(d\vt^c\wedge\vt_c)
\Big)+\nonumber\\
&&\rho_3\Big(
-2\vt_b \wedge d*( \vt_a \wedge d \vt^b )+
2d\vt_b\wedge*(\vt_a\wedge d\vt^b)+\nonumber\\
&&
e_a\hook\Big(\vt_c\wedge d\vt^b\wedge*(d\vt^c\wedge\vt_b)\Big)-
2(e_a\hook d\vt^b)\wedge\vt_c\wedge*(d\vt^c\wedge\vt_b )
\Big)=0.
\err
The general ( ``diagonal'')  
spherical-symmetric static  solution to the field equation
(\ref{3.6}) for all possible values of $\rho_i$ is derived \cite{it5}. 
It turns out  that $\rho_1=0$ is a necessary and sufficient condition 
to have a solution with Newtonian behavior at infinity. 
The coframe solution  in this case is unique and 
yields via (\ref{3.1}) the Schwarzschild metric. 
In such a way by rejecting the pure Yang-Mills-type 
term (\ref{3.3}) the model turns out to be  a viable model for gravity.\\ 
Few remarks on the analytic structure of the field equation (\ref{3.6}) 
are now in order. \\
On one hand, the coframe field is a complex of $16$ independent variables 
while the symmetric metric tensor field has only 
$10$ independent components. 
The remaining 6 components are related to the spinorial properties of the field. 
An additional {\it local} $SO(1,3)$ {\it invariance}, 
which appears in the case 
\begin{equation}\label{3.9}   
\rho_1=0, \qquad \rho_2+2\rho_3=0, 
\end {equation} 
 restricts the set of 16 independent variables to a subset of $10$ variables.  
This subset is in one to one correspondence with 10  independent components 
of the metric. \\
On the other hand 
the field equation (\ref{3.6}) is a system of 16 independent equations. 
This system is   
reduced to two covariant systems - a symmetric tensorial sub-system of 
10 independent equations and an antisymmetric tensorial sub-system of 6 
independent equations.  
In the case (\ref{3.6}) (and only in this case) the antisymmetric equation 
vanishes identically and the system  is restricted to a system of 10 
independent equations for 10 independent variables. 
Therefore the local $SO(1,3)$ invariant 
coframe  structure coincides with the metric structure. 
The model with parameters (\ref{3.9})  is referred to as the 
{\it teleparallel equivalent of gravity}.  
This local invariant construction, in fact, is not an alternative model of gravity 
but merely an alternative coframe reformulation of the standard (tensorial) GR. 
Namely, the Lagrangian (\ref{3.2}) with the 
parameters determined by (\ref{3.9}) coincides with 
the Hilbert-Einstein Lagrangian (up to total derivative terms) \cite{H-S}. \\
In the general case, when the relations  (\ref{3.9}) do not hold, 
the field equation (\ref{3.6}) is a well defined covariant system 
of 16 independent equations for 16 independent variables. \\
Certainly the most interesting case is 
\begin{equation}\label{3.9a}   
\rho_1=0, \qquad \rho_2, \rho_3 \quad {\textrm - \ arbitrary}
\end {equation}
For these values of parameters the Lagrangian  (\ref{3.2}) and, consequently, 
the field equations (\ref{3.6}) describe a 1-parametric 
family of models with a unique ``diagonal'' 
spherical-symmetric solution  which yields the Schwarzschild metric. 
Hence all the models of the family conform to  the observation data at 
least for the three classical tests of GR. 
Thus the family of models (\ref{3.9a}) provides a viable alternative to the 
standard GR. 
\section{Coframe field}             
\subsection{The compact form of the  Lagrangian}         
We study an  even  smooth coframe field $\vt^a$   defined on a differential  
$4D$-manifold $M$. 
Our goal is to derive a conserved current expression 
for this coframe  field  in a  set 
of models parameterized by the constants $\rho_i$. 
Although there are good physical reasons for rejecting the pure 
Yang-Mills term in the 
Lagrangian by taking $\rho_1=0$, the general case is not more difficult for treatment, so  
we will consider the complete set of teleparallel models (\ref{3.2}) with arbitrary 
values of parameters.\\
The standard computations of the variation of a Lagrangian defined on 
a teleparallel manifold are rather complicated \cite{T-W}, \cite{Hehl98}. 
It is because one needs to vary not only  the coframe $\vt^a$ 
itself, but also the the dual frame $e_a$ and even the  
Hodge dual operator $*$, 
that  depends on the pseudo-orthonormal coframe implicitly.\\
In order to avoid these technical problems we will rewrite  
the total Lagrangian (\ref{3.2}) in a 
compact form which will be useful for the  
variation procedure.\\
Consider the exterior differentials of the basis 1-forms $d\vt^a$ and 
introduce the $C$-coefficients of their  expansion in the basis of 
even 2-forms $\vt^{ab}$ 
(here and later  the abbreviation 
$\vt^{ab\cdots}=\vt^a\wedge \vt^b\wedge \cdots$ is used)
\begin{equation}\label{4.2} 
d\vt^a=\vt^a_{\b,\a}dx^\a\wedge dx^\b:=\frac 12 {C^a}_{bc}\vt^{bc}.
\end{equation}
By definition, the coefficients ${C^a}_{bc}$ are  antisymmetric: 
${C^a}_{bc}=-{C^a}_{cb}.$ 
Their explicit expression is derived straightforward from the definition 
(\ref{4.2})
\begin{equation}\label{4.3} 
{C^a}_{bc}:=e_c\hook(e_b\hook d\vt^a).
\end{equation}
In terms of the $C$-coefficients the independent parts of 
the Lagrangian (\ref{3.2})  are 
\brr\label{4.4}   
{}^{(1)}L&=& \frac 12 C_{abc}C^{abc}*1,\nonumber\\
{}^{(2)}L&=& \frac 12 C_{abc}\Big(C^{abc}+C^{bca}+C^{cab}\Big)*1,\nonumber\\
{}^{(3)}L&=&\frac 12 \Big(C_{abc}C^{abc}-2{C^a}_{ac}{C_b}^{bc}\Big)*1.   
\err
Note that the form (\ref{4.4}) is useful for a proof of the 
completeness of the set of 
quadratic invariants \cite{i-k}. 
It is enough to consider all the possible combinations of the indices. 
Thus a linear combination of the Lagrangians (\ref{4.4}) is the most general 
quadratic coframe Lagrangian.  \\
Using  (\ref{4.4}) we rewrite the  coframe Lagrangian 
in a compact form
\begin{equation}\label{4.5}  
L=\frac 1{4} C_{abc}C_{ijk}\lambda^{abcijk}*1,
\end{equation}
where the  constant symbols
\brr\label{4.6}  
\lambda^{abcijk}&:=&(\rho_1+\rho_2+\rho_3)\eta^{ai}\eta^{bj}\eta^{ck}+
\rho_2(\eta^{aj}\eta^{bk}\eta^{ci}+\eta^{ak}\eta^{bi}\eta^{cj})\nonumber \\
&&-2\rho_3\eta^{ac}\eta^{ik}\eta^{bj}
\err
are introduced.  
It can be checked, by straightforward calculation, that these $\lambda$-symbols 
are invariant under a 
transposition of the triplets of indices: 
\begin{equation}\label{4.7}
\lambda^{abcijk}=\lambda^{ijkabc}.
\end{equation}
We also  introduce an abbreviated notation 
 \begin{equation}\label{4.8}
F^{abc}:=\lambda^{abcijk}C_{ijk}.
\end{equation}
The total Lagrangian (\ref{3.2}) reads now as
\begin{equation}\label{4.10}
L=\frac 1{4} C_{abc}F^{abc}*1.
\end{equation}
This form of the Lagrangian will be used in the consequence 
for the variation procedure. 
The Lagrangian (\ref{4.10}) can also be rewritten in a component free 
notations. \\
Define one-indexed 2-forms: a strength form 
 \begin{equation}\label{4.16}
\C^a:=\frac 12 C^{abc}\vt_{bc}=d\vt^a.
\end{equation}
and a conjugate strength form $\F^a:=\frac 12 F^{abc}\vt_{bc}$ 
 \begin{equation}\label{4.16a}
\F^a=(\rho_1+\rho_3)\C^a+
\rho_2e^a\hook(\vt^m\wedge\C_m)-\rho_3\vt^a\wedge(e_m\hook \C^m)
\end{equation}
The 2-form $\F^a$ can be also represented via the irreducible (under the 
Lorentz group) decomposition of the 2-form $\C^a$ (see \cite{Hehl98}, \cite{McCrea}). 
Write 
\begin{equation}\label{4.17a}
\C^a={}^{(1)}\C^a+{}^{(2)}\C^a+{}^{(3)}\C^a,
\end{equation}
where
\brr\label{4.17b}
{}^{(1)}\C^a&=&\C^a-{}^{(2)}\C^a-{}^{(3)}\C^a,\nonumber\\
{}^{(2)}\C^a&=&\frac 13 \vt^a\wedge(e_m\hook \C^m),\nonumber\\
{}^{(3)}\C^a&=&\frac 13 e^a\hook(\vt_m\wedge \C^m).
\err
Substitute (\ref{4.17b}) into (\ref{4.16a}) to obtain 
\begin{equation}\label{4.17c}
\F^a=(\rho_1+\rho_3){}^{(1)}\C^a+(\rho_1-2\rho_3){}^{(2)}\C^a+
(\rho_1+3\rho_2+\rho_3){}^{(3)}\C^a.
\end{equation}
The coefficients in (\ref{4.17c}) coincide with those calculated in 
\cite{Hehl98}. \\ 
The 2-forms  $\C^a$ and $\F^a$ do not depend on a choice of 
a coordinate system. They  change as 
vectors by global $SO(1,3)$ transformations of the coframe. 
Using (\ref{4.16}) the coframe Lagrangian can be rewritten  as
 \begin{equation}\label{4.17}
L=\frac 1{2} \C_a\wedge *\F^a
 \end{equation}
Observe that the Lagrangian (\ref{4.17}) is of the same form as the 
standard electromagnetic Lagrangian 
$L=\frac 1{2} F\wedge*F.$
However, the teleparallel Lagrangian involves the vector valued 2-forms of the 
field strength, while the  electromagnetic Lagrangian is constructed of the 
the scalar valued 2-forms. 
 \subsection{Variation of the Lagrangian}                
The Lagrangian (\ref{4.17}) depends on the coframe field $\vt^a$ and on its first 
order derivatives only. 
Thus the first order variation formalism guarantee the corresponding 
Euler-Lagrange equation to be of at most second order. 
Consider the variation of the coframe Lagrangian,  taken in the 
component-wise form (\ref{4.10}),  relative to small independent variations 
of the 1-forms $\vt^a$. 
The $\lambda$-symbols (\ref{4.6}) are constants and obey the 
symmetry property (\ref{4.7}). 
Thus 
\begin{equation}
C_{abc}\d F^{abc}=C_{abc}\lambda^{abcijk}\d C_{ijk}=\d C_{abc} F^{abc}
\end{equation}
Consequently 
the variation of the Lagrangian  (\ref{4.10}) takes the form 
\begin{equation}\label{4.11}  
\d L=\frac 1{2}\d C_{abc}F^{abc}*1-L*\d(*1).
\end{equation}
The variation of the volume element is 
\brn   
\d(*1)&=&-\d(\vt^{0123})=-\d\vt^0\wedge \vt^{123}-\cdots=
-\d\vt^0\wedge *\vt^0-\cdots \nonumber  \\
&=&
\d \vt^m\wedge*\vt_m.
\ern   
Thus
\begin{equation}\label{4.12}
L*\d(*1)=(\d\vt^m\wedge *\vt_m)*L=-\d\vt^m\wedge(e_m\hook L).
\end{equation}
As for the variation of the $C$-coefficients, we calculate them by equating  
the variations  of the two sides of the equation (\ref{4.2})
\begin{equation}
\d d\vt_{a}=\frac 12 \d C_{amn}\vt^{mn}+C_{amn}\d\vt^{m}\wedge\vt^n.
\end{equation}
Use the formulas (\ref{A.12}) and (\ref{A.15}) to derive  
\brn
\d d\vt_{a}\wedge*\vt_{bc}&=&
\frac 12 \d C_{amn}\vt^{mn}\wedge*\vt_{bc}+
C_{amn}\d\vt^{m}\wedge\vt^n\wedge*\vt_{bc}\\
&=&-\frac 12 \d C_{amn}\vt^{m}\wedge *(e^n\hook\vt_{bc})-
C_{amn}\d\vt^{m}\wedge*(e^n\hook\vt_{bc})\\
&=&\d C_{abc}*1-2\d\vt^{m}\wedge C_{am[b}*\vt_{c]}.
\ern
Therefore 
\begin{equation}\label{4.13}
\d C_{abc}*1=\d (d\vt_a)\wedge *\vt_{bc}
+2\d\vt^m \wedge C_{am[b}*\vt_{c]}.
\end{equation}
After substituting (\ref{4.12}--\ref{4.13}) into (\ref{4.11})
the variation of the Lagrangian  takes the form
\brn
\d L&=&
\frac 1{2} F^{abc}\Big(\d (d\vt_{a})\wedge*\vt_{bc}+2\d\vt^m\wedge C_{am[b}*\vt_{c]}\Big)
+\d \vt^m \wedge (e_m\hook L).
\ern
Extract total derivatives  to obtain
 \brr\label{4.15}
\d L &=&\frac 1{2}
\d\vt_{m}\wedge\Big( d(*F^{mbc}\vt_{bc})+
2F^{abc}C_{am[b}*\vt_{c]}
+2 e_m\hook L\Big)\nonumber\\
&&+\frac 1{2}d\Big(\d\vt_{a}\wedge*F^{abc}\vt_{bc}\Big).
\err
The variation relation (\ref{4.15}) will play a basic role in the sequel. 
Let us rewrite it in a compact form by using the 2-forms (\ref{4.16}) and 
(\ref{4.16a}). 
The terms of the form $F\cdot C$  can be rewritten as 
\brn
&&F^{abc}C_{am[b}*\vt_{c]}=(F^{abc}-F^{acb})C_{am[b}*\vt_{c]}
\nonumber\\
&&\qquad 
= C_{amb}*(e^b\hook \F^a)=-(e_m\hook \C_a)\wedge *\F^a.
\ern
Hence, (\ref{4.15}) takes the form 
 \brr\label{4.18}
\d L&=& \d\vt^m\wedge \Big(d(*\F_m)-(e_m\hook \C_a)\wedge *\F^a
+e_m\hook L\Big)+d(\d\vt^m\wedge \F_m).
\err
Collect now the quadratic terms  into a differential 3-form 
 \begin{equation}\label{4.20}
\T_m:=(e_m\hook \C_a)\wedge *\F^a - e_m\hook L.
 \end{equation}
Consequently,   the variational relation (\ref{4.15}) 
results in the final form
\begin{equation}\label{4.22}
\d L= \d\vt^m\wedge \Big(d*\F_m-\T_m\Big)+
d(\d\vt^m\wedge \F_m).
\end{equation}
 \subsection{The field equations}                 
We are ready now to write down the field equations.
Consider independent free variations of a  coframe field vanishing at infinity 
(or at the boundary of the manifold $\partial M$).  
The variational relation (\ref{4.22}) yields {\it the coframe field equation}   
\begin{equation}\label{4.24} 
d*\F^m=\T^m.
\end{equation}
Note that this is the same equation as (\ref{3.6}) because it was 
obtained from the same Lagrangian by the same free variations of the  coframe. 
The equivalence of the forms is shown in the  Appendix B. \\
Observe that the structure of coframe field equation is formally similar to the 
structure of the standard  electromagnetic field equation $d*F=J$. 
Namely, the left hand side  of both equations 
is the exterior derivative of the dual strength 
field while the right hand side is an odd 3-form. 
Thus the 3-forms $\T^m$ serves as a source for the strength field  $\F^m$, 
as well as the 3-form of electromagnetic current $J$ is a source for the 
electromagnetic field.\\
There are, however, important distinctions:\\
i) The coframe field current $\T_m$ is a vector-valued 3-form while 
the electromagnetic current $J$ is scalar-valued. \\
ii) The field equation (\ref{4.24}) is nonlinear. \\
iii) The electromagnetic current $J$   depends on an exterior material field, 
while the coframe current $\T^m$ is interior (depends on the coframe itself). \\ 
The exterior derivation  of the 
field equation (\ref{4.24}) yields the conservation law
\begin{equation}\label{cons-I}
d\T_m=0.
\end{equation}
Note, that this equation obeys all symmetries of the Lagrangian. 
It is diffeomorphism invariant and global $SO(1,3)$ covariant.  
Thus we obtain a conserved total 3-form (\ref{4.20}) which is 
constructed from the first order derivatives of the field variables (coframe). 
It is local and covariant.  
The 3-form $\T_m$ is our  candidate for the coframe energy-momentum current. 
\subsection{Conserved currents}                       
The current $\T_m$ is obtained 
directly, i.e., by separation of the terms in  the field equation.   
In order to identify the nature of this conserved 3-form  
we have to answer the question: 
{\bf{\it What symmetry this conserved current can be associated with?}}\\
Return to the variational relation (\ref{4.22}). 
On shell, for the fields satisfying the field equations (\ref{4.24}), 
it takes the form
\begin{equation}\label{4.26}   
\d L=d (\d\vt^a\wedge *\F_a).
\end{equation}
Consider the variations of the coframe field produced by the 
Lie derivative taken relative to a smooth vector field $X$, i.e., 
\begin{equation}\label{4.28a} 
\d\vt^a=\L_X\vt^a=d(X\hook\vt^a)+X\hook d\vt^a.
\end{equation}
The Lagrangian (\ref{4.10}) is a diffeomorphic invariant, 
hence it's  variation is produced by the Lie  derivative taken relative 
to the same vector field $X$, i.e., 
\begin{equation}\label{4.28b} 
\d L=\L_X L=d(X\hook L). 
\end{equation}
Thus the relation (\ref{4.26}) takes a 
form of a conservation law $d\Theta(X)$ for the Nether 3-form
\begin{equation}\label{4.29} 
\Theta(X):=\Big(d(X\hook\vt^a)+X\hook \C^a \Big)\wedge *\F_a-X\hook L.
\end{equation}
This  quantity includes the derivatives of an   
arbitrary vector field $X$. 
Such a non-algebraic dependence  of the conserved current is an obstacle 
for definition of an energy-momentum tensor. 
This problem is solved 
merely by using the canonical form of the current. 
Let us  take $X=e_a$. The first term of  (\ref{4.29})  vanishes  
identically.  Thus  
\begin{equation}\label{4.30} 
\Theta(e_m)=(e_m\hook \C^a )\wedge *\F_a-e_m\hook L.
\end{equation}
Observe that  the right hand side of the equation (\ref{4.30}) is 
exactly the same expression as
 the source term of the field equation (\ref{4.24}): 
\begin{equation}\label{4.30a} 
\Theta(e_m)=\T_m
\end{equation}
Thus the conserved current $\T_m$ defined in (\ref{4.20}) 
is associated with the diffeomorphism invariance of the Lagrangian. 
Consequently the vector-valued 3-form (\ref{4.20}) represents 
the energy-momentum current of the coframe field. 
\subsection{Noether charge}                       
Let us look for  an additional information incorporated 
in the conserved current (\ref{4.29}).
Extract the total derivative to obtain
 \begin{equation}\label{4.31aa} 
\Theta(X)=d\Big((X\hook\vt^a)*\F_a\Big)-(X\hook\vt^a) (d*\F_a-\T_a) 
\end{equation}
Thus, up to the field equation (\ref{4.24}), the current $\T(X)$ represents  
a total derivative of a certain 2-form 
$\Theta(X)=dQ(X)$. 
This result is a special case of a general proposition due 
to Wald \cite{Wald2}   for  a diffeomorphic invariant Lagrangians. 
The  2-form  
\begin{equation}\label{4.31} 
Q(X)=(X\hook \vt^a) *\F_a.
\end{equation}
is  referred to as the {\it Noether charge for the coframe field}. 
Consider $X=e_a$ and denote $Q_a:=Q(e_a)$. 
From (\ref{4.31}) we obtain that this canonical Noether charge
of the coframe field coincides with the dual of the conjugate strength. 
\begin{equation}\label{4.32} 
Q_a=Q(e_a)=*\F_a.
\end{equation} 
In this way the 2-form $\F_a$, which was used above only as a technical 
device for expressing the  equations in a compact form, obtained now 
a meaningful description. 
Note, that the Noether charge plays an important role in Wald's treatment 
of the black hole entropy \cite{Wald2}.

\subsection{Energy-momentum tensor}                       
In this section we construct an expressions for the energy-momentum tensor 
for the coframe field.   
Let us first introduce the notion of the energy-momentum tensor 
by the differential-form formalism. 
We are looking for a second rank tensor field of a type $(0,2)$.
Such  a tensor can always be treated as a  bilinear map
$
T: \ {\mathcal X(M)}\times{\mathcal X(M)} \to {\mathcal F(M)},
$ 
where ${\mathcal F(M)}$ is the algebra of $C^\infty$-functions on $M$
while ${\mathcal X(M)}$ is the ${\mathcal F(M)}$-module of vector fields
on $M$. 
The unique way to construct a scalar from a 3-form and a vector is 
is to take the Hodge dual of the 3-form and to 
contract the result by the vector. 
Consequently, we define the energy-momentum tensor as 
\begin{equation}\label{2.27}  
T(X,Y):=Y\hook *\T(X).
\end{equation}
Observe that this quantity is a tensor if and only if the 3-form current
$\T$ depends  linearly (algebraic) on the vector field $X$.
Certainly, $T(X,Y)$ is not symmetric in general. 
The antisymmetric part of the energy-momentum tensor is known from 
the Poincar{\'e} gauge theory \cite{Hehl4} to represent the spinorial current of the 
field. \\
The canonical form of the energy-momentum $T_{ab}:=T(e_a,e_b)$ tensor is 
\begin{equation}\label{2.28}    
T_{ab}=e_b\hook *\T_a.
\end{equation}
Another useful form of this tensor can be obtained from (\ref{2.28}) 
by applying the rule (\ref{A.15}) 
\begin{equation}\label{2.29}
T_{ab}=*(\T_a\wedge \vt_b).
\end{equation}
The familiar procedure of rising the indices by the Lorentz metric $\eta^{ab}$ 
produces two tensors of a type $(1,1)$
\begin{equation}\label{2.30}
{T_a}^b=*(\T_a\wedge \vt^b),
\quad {\textrm {and}} \quad
{T^a}_{b}=*(\T^a\wedge \vt_b),
\end{equation}
which are different, in general.  
By applying the rule (\ref{A.9}) the first relation of (\ref{2.30}) 
is converted into
\begin{equation}\label{2.32}
\T_a={T_a}^b*\vt_b.
\end{equation}
Thus, the components of the energy-momentum tensor are regarded as  
the coefficients of the current $\T_a$ in the dual basis
$*\vt^a$ of the vector space $\Omega^3$ of odd 3-forms.\\
In order to show that  (\ref{2.32}) conforms with the intuitive notion 
of the energy-momentum tensor
let us represent  on the flat manifold the 3-form conservation law 
as a  tensorial conservation law. 
Take a closed  coframe $d\vt^a=0$, thus $d*\vt_b=0$. 
From (\ref{2.32}) we derive 
\brn
d\T_a=d{T_a}^b\wedge *\vt_b  =-{{T_a}^b}_{,b}*1.
\ern
Hence the differential-form conservation law $d\T_a=0$
is equivalent to the tensorial conservation law ${{T_a}^b}_{,b}=0$.\\ 
Apply now the definition (\ref{2.28}) to the conserved current
(\ref{4.20}) for the coframe field.
The  energy-momentum tensor $T_{mn}=e_n\hook*\T_m$ 
is derived in the form 
\begin{equation}\label{4.39}
T_{mn}=e_n\hook*\Big((e_m\hook\C_a)\wedge*\F^a-
\frac 12 e_m\hook(\C_a\wedge*\F^a)\Big).
\end{equation}
Using (\ref{A.15}) we rewrite the first term in (\ref{4.39}) as 
\brn
&&e_n\hook*\Big((e_m\hook\C_a)\wedge*\F^a\Big)=
-*\Big((e_m\hook\C_a)\wedge*(e_n\hook\F^a)\Big).
\ern
As for the second term in (\ref{4.39}) it takes the form 
\brn
&&-\frac 12 e_n\hook*\Big(e_m\hook(\C_a\wedge*\F^a)\Big)=
\frac 12\eta_{mn}*(\C_a\wedge*\F^a).
\ern
Consequently the energy-momentum tensor for the coframe field is  
\begin{equation}\label{4.40}
T_{mn}=-*\Big((e_m\hook\C_a)\wedge*(e_n\hook\F^a)\Big)+
\frac 12\eta_{mn}*(\C_a\wedge*\F^a).
\end{equation}
Observe that this expression is formally similar to the familiar 
expression for the energy momentum tensor of the Maxwell 
electromagnetic field: 
\begin{equation}\label{2.34a}       
T_{mn}=-*\Big((e_m\hook F)\wedge *(e_n\hook F)\Big)+
\frac 12\eta_{mn}*(F\wedge *F).
\end{equation}
The form (\ref{2.34a}) is no more than an expression of the 
electromagnetic energy-momentum tensor in arbitrary frame. 
In a specific coordinate chart $\{x^\mu\}$ it is enough to take the coordinate 
basis vectors $e_a=\partial_\a$ and consider 
$T_{\a\b}:={}^{(e)}T(\partial_\a,\partial_\b)$ 
 to obtain the familiar  expression 
 \begin{equation}\label{2.35}  
T_{\a\b}=
 -F_{\a\m}{F_\b}^\m+\frac 14 \eta_{\a\b}F_{\m\n}F^{\m\n}.
\end{equation}
The electromagnetic energy-momentum  tensor is obviously traceless.  
The same property holds also for the coframe field tensor. \\
{\bf Proposition}
{\it For all teleparallel models described by the Lagrangian (\ref{3.2}), i.e., for all values 
of the parameters $\rho_i$,  
the energy-momentum tensor defined by 
(\ref{4.40}) is traceless.}\\
{\bf Proof}
Compute the trace ${T^m}_m=T_{mn}\eta^{mn}$ of (\ref{4.40}): 
\brn
&&{T^m}_m=-*\Big((e_m\hook\C_a)\wedge*(e^m\hook\F^a)\Big)+2*(\C_a\wedge*\F^a)\\
&&=*\Big((e_m\hook\C_a)\wedge*^2(\vt^m\wedge*\F^a)\Big)+2*(\C_a\wedge*\F^a)\\
&&=-*\Big(\vt^m\wedge(e_m\hook\C_a)\wedge*\F^a\Big)+2*(\C_a\wedge*\F^a)=0
\ern
In the latter equality the relation  (\ref{A.9}) was used. \\
It is well known that the traceless of the energy-momentum tensor is 
associated  with the scale invariance of the Lagrangian. 
The rigid ($\lambda$ is a constant) scale transformation
$x^i\to \lambda x^i$, is considered acting on a 
material field as $\phi\to \lambda^d\phi$, where $d$ is the 
dimension of the field. 
The transformation does not act, however, on the components 
of the metric tensor and 
on the frame (coframe) components. 
It is convenient to shift the change on the metric and on the frame 
(coframe)  components: 
$g_{\mu\nu}\to \lambda^2g_{\mu\nu}$ , 
$ {\vt^a}_{\mu}\to \lambda{\vt^a}_{\mu}$, and 
$ {e_a}^{\mu}\to \lambda^{-1}{\vt_a}^{\mu}$
with no change of coordinates. 
In the coordinate free approach the difference between two approaches is neglected and the transformation is 
\begin{equation}\label{4-42b}
g\to\lambda^2 g,\qquad \vt^a\to \lambda\vt^a, \qquad \textrm{and} \qquad e_a\to \lambda^{-1}e_a
\end{equation}
The transformation law of the teleparallel Lagrangian is simple to obtain from the 
component-wise form (\ref{4.4}). 
Under the transformation (\ref{4-42b}) the volume element changes as 
$*1\to \lambda^4 *1$. 
As for the $C$-coefficients, they transform due to (\ref{4.3}) as ${C^a}_{bc}\to\lambda^{-1}{C^a}_{bc} $. 
Consequently, by (\ref{3.3}),  the transformation law of the Lagrangian 4-form is $L\to\lambda^2 L$, 
which is the same as for the Hilbert-Einstein Lagrangian 
$L_{HE}=R\sqrt{-g}d^4x\to\lambda^2 L_{HE}$. 
After rescaling the Planck length  the scale invariance is reinstated. 
Hence, for the pure teleparallel model the energy-momentum tensor have to 
be traceless in accordance with the proposition above. 
 \subsection{The field equation for a general system}           
The coframe field equation have been derived for a pure coframe field. 
Consider now a general minimally coupled system of  
a coframe field $\vt^a$ and a material field $\psi$. 
The material field can be  a differential form of an arbitrary degree 
 and can carry arbitrary number of exterior and interior indices. 
Take the total Lagrangian of the system to be of the form ($\ell=$ Planck
length)
\begin{equation}\label{4-41}
L=\frac 1{\ell^2}L^{cof}(\vt^a,d\vt^a)+L^{mat}(\vt^a,\psi, d\psi),
\end{equation}
where the coframe Lagrangian $L^{cof}$, defined by (\ref{3.2}), 
is of dimension length square,  
while the material Lagrangian $L^{mat}$ is dimensionless. \\
The minimal coupling means here the absence of coframe derivatives 
in the material Lagrangian.  
Take the variation of (\ref{4-41}) relative to the coframe field $\vt^a$ 
to obtain 
\begin{equation}\label{4-42}
\d L=\frac 1{\ell^2}\d\vt^a\wedge\Big(d*\F_a-\T^{cof}_a-{\ell^2}\T^{mat}_a\Big),
\end{equation}
where the 3-form of coframe current is defined by (\ref{4.24}). 
The 3-form of material current  is defined via the variation 
derivative of the material Lagrangian taken relative to the coframe field $\vt^a$:
\begin{equation}\label{4-42aa}
\T^{mat}_a:=-\frac {\d}{\d\vt^a}L^{mat}.
\end{equation}
Introduce the total current of the system 
$\T^{tot}_a=\T^{cof}_a+{\ell^2}\T^{mat}_a$,
which is of dimension length (mass). 
Consequently, the field equation for the general system (\ref{4-41}) takes the 
form 
\begin{equation}\label{4-43}
d*\F_a=\T^{tot}_a.
\end{equation}
Using the energy-momentum tensor 
(\ref{2.32})  this equation can be rewritten in a tensorial form 
\begin{equation}\label{4-44}
e_b\hook *d*\F_a=T^{tot}_{ab},
\end{equation}
or equivalently 
\begin{equation}\label{4-45}
\vt_b\wedge d*\F_a=T^{tot}_{ab}*1.
\end{equation}
The conservation law for the total current $d\T_a=0$ is a straightforward 
consequence of the field equation (\ref{4-43}). \\
The form (\ref{4-43}) of the field equation looks like 
 the Maxwell field equation for the 
electromagnetic field $d*F=J$. 
Observe, however, an important difference. \\
The source term in the right hand side of the electromagnetic field equation 
depends only on external fields. 
In the absence of the external sources $J=0$, 
the electromagnetic strength $*F$ is a closed form. 
As a consequence, its cohomology class interpreted as a charge of the 
source. 
The electromagnetic field itself is uncharged.\\
As for the coframe field strength $\F^a$ its source depends on the 
coframe and of its first order derivatives. 
Consequently, the 2-form $*\F^a$ is not closed even in absence 
of the external sources. 
Hence the gravitational field is massive (charged) itself. \\ 
On the other hand the tensorial form (\ref{4-44}) of the teleparallel 
field equation is similar to the 
Einstein field equation for the metric tensor
$
G_{ab}=8\pi T^{mat}_{ab}.
$
Indeed, the left hand side in both equations are pure geometric 
quantities. 
Again, the source terms in the field equations  
are  different. 
The source of the Einstein gravity is the energy-momentum tensor  
only of the  materials fields. 
The conservation of this tensor is a consequence of the field equation. 
Thus even if  some meaningful conserved energy-momentum current for 
the metric field existed  
it would have been conserved regardless of  the material field current. 
Consequently, any redistribution of the energy-momentum current between the 
material 
and gravitational fields is  forbidden in the framework of the traditional 
Einstein gravity.  \\
As for the coframe field equation, the total energy-momentum current plays 
a role of the source of the field. 
Consequently the coframe field is completely ``self-interacted'' - the 
energy-momentum current of the coframe field produces an additional field. 
The conserved current of the coframe-material system  is the total 
energy-momentum current, not only the material current. 
Thus in the framework of general teleparallel construction 
the redistribution of the current between  the material field and the 
coframe field is, in principle, possible. 
\section{Teleparallel equivalent of GR}
The gravitational energy-momentum problem attracted recently a 
considerable interest in the framework of the teleparallel equivalent of GR 
model (denote it by ${\textrm {GR}}_{||}$) \cite{}. 
As it was mentioned, this model corresponds to a special choice (\ref{3.9}) 
of free parameters of the general teleparallel model described above. 
Let us start with a comparison between the differential form approach 
and the tensorial approach used in the ${\textrm {GR}}_{||}$. \\
1) The basic dynamical variable of the ${\textrm {GR}}_{||}$ is the frame (tetrad) 
field ${h^a}_\mu$, where the Greek index is related to the  
coordinates  while 
the Latin index  denotes the corresponding vector in the frame.  
Due to the canonical duality between 1-forms and vectors this 
set of variables is equivalent to the components of the coframe field 
taken in coordinate basis $\vt^a={\vt^a}_{\mu}dx^{\mu}$. \\
2) The gravitational field strength of the ${\textrm {GR}}_{||}$ is the torsion 
tensor ${T^\rho}_{\mu\nu}={h_a}^\rho\partial_{[\nu}{h^a}_{\mu]}$. 
This object is in one to one correspondence with  
the coefficients of the 2-form $\C^a=d\vt^a$ taken in  a coordinate basis. 
The second field strength tensor of the ${\textrm {GR}}_{||}$, the contorsion tensor, 
is defined as a linear combination of the torsion tensor with the coefficients 
depend on the metric tensor $g^{\mu\nu}$. 
It corresponds  to the linear combinations of the components ${C^a}_{bc}$, 
used above.\\
3) The Lagrangian of the ${\textrm {GR}}_{||}$, its field equations and conserved current 
are constructed from the torsion and contorsion tensors and reinstated 
from the formulas above in a special case of parameters  (\ref{3.9}) 
being written in a coordinate basis. \\
Thus the two techniques: the tensorial representation of the ${\textrm {GR}}_{||}$ and the 
differential form approach used above are principally equivalent. 
It should be noted, however, that the very fact of treating the gravitational 
strength 
as  the antisymmetric tensor  of torsion shows that the differential forms  
approach is a  more appropriative mathematical device here. \\
Observe now the principal features of the ${\textrm {GR}}_{||}$ model. 
Certainly in the framework of general coframe model 
the construction do not depends on the specific values 
of the parameters $\rho_i$. 
The ${\textrm {GR}}_{||}$ Lagrangian is reinstated from the 
general Lagrangian  (\ref{3.2}) 
merely by inserting the specific values of the coefficients. 
Also the field equation and the conserved current do not depend on 
a choice of the parameters. 
Thus, it seems that the ${\textrm {GR}}_{||}$-model can 
be considered as a simple limit 
$\rho_1\to 0, \ \rho_2+2\rho_3\to 0$ of the general teleparallel construction.  
A more detailed analyses shows, however, that it is not a case. \\
Even the limit $\rho_1\to 0$ is not trivial. 
Indeed, the typical form of the spherical-symmetric solution in 
the general model \cite{it5} is 
 $\vt^a=(r/r_0)^\alpha dx^a$, where $\alpha$ depends of the coefficients 
$\rho_i$. 
The Schwarzschild coframe is a special solution, which appears in the 
case $\rho_1=0$ only. 
Certainly, the typical solution does not approach the Schwarzschild
coframe in the limit.  
In fact the general free parametric model has a non-continuous dependence on the 
parameter $\rho_1$. \\
The second limit $\rho_2+2\rho_3\to 0$ produces an additional 
degeneration of the general coframe construction. 
The Lagrangian  (\ref{3.2}) obtains in this case a higher symmetry. 
This is a local Lorentz invariance of the coframe field. 
Due to the known theorem of the variational calculus this invariance 
appears also on the field equation level. 
However, the separation of the field equation to the total derivative term 
$d*\F^a$ and the conserved current term $\T^a$ is not local Lorentz 
invariant. 
As a result in the  ${\textrm {GR}}_{||}$ limit the notion of the 
conserved current and 
of the gravitational energy-momentum tensor has not an invariant sense. 
Although, this object is invariant under diffeomorphism transformations 
of the manifold, it is not invariant under local $SO(1,3)$ transformations of the 
coframe.  
Thus the conserved current of ${\textrm {GR}}_{||}$ inherited from the general 
free parametric model is no more than a 
type of a pseudo-tensor. 
Its connection to the M{\o}ller pseudo-tensor is shown in \cite{Per}. 
\section{Conclusions and discussion}
We considered a general 3-parametric teleparallel model in a 
coframe representation. 
The field equations and the conserved current (vector-valued 3-form) are 
derived via the variation procedure. 
By using the Noether technique the current is shown to be produce by the 
invariance of the Lagrangian under diffeomorfism transformations of 
the coframe.  
Consequently it is the energy-momentum current. 
The energy momentum-tensor of the coframe field is constructed. 
We considered a minimal coupling system of a coframe and a material field. 
It is shown that the total energy-momentum current of the system plays a role 
of  the source of the coframe field strength.   
The total current is conserved, which yields a possibility of redistributing the 
of the energy between the coframe and the material field. 
Such effect is forbidden in the framework of the standard GR. 
A special case of the teleparallel equivalent of GR is discussed. 
This model is derived from the general  construction by a specification of the 
free parameters. 
However, the conserved current of this equivalent of GR has not an 
invariant sense. 
It is because of localization of the Lorentz symmetry. \\
The result is: The standard GR has in the parametric space 
a neighborhood of viable models with the 
same Schwarzschild solutions. 
This models however have a better Lagrangian behavior and produce an invariant 
energy-momentum tensor.\\
The study of general teleparallel models can be interesting from two points of view: \\
1) As a family of {\it viable alternative models} of gravity. For this the parameters 
should be taken as $\rho_1=0$ and $\lambda=\rho_2+2\rho_3\ne 0$. 
Certainly this teleparallel construction is different from the Einstein 
theory in the treatment of the  axial symmetric spaces. 
The exact solution of such type can give a good indication of viability of 
this alternative model. 
Another line is to study alternative models of coupling of gravity with material fields. 
It is also important to search for a bound on the parameter $\lambda$. \\ 
2) The coframe approach can serve as an {\it alternative formulation of the standard GR}. 
This formalism can be helpful for the treatment of the energy-momentum problem 
of GR in the integral (quasi-local) aspect. 
An appropriative defined integral of the non-local invariant teleparallel current 
hoped to preserve the local invariance. 
The examples of such type behavior are well known from the electromagnetism theory. 

\section*{Acknowledgments}
I am deeply grateful to F.W. Hehl and to S. Kaniel  for their critical   
reading and valuable suggestions. 
\appendix

\section{Basic notations and definitions}
Let us list our basic conventions. 
We consider an $n$-dimensional differential manifold $M$
of signature
\begin{equation}\label{A.1}
\eta_{ab}=diag(-1,+1,\cdots,+1).
\end{equation}
Let the manifold $M$ will be endowed with a smooth coframe field
(1-forms) 
\begin{equation}\label{A.2}
\{\vt^a(x), \ a=0,\cdots,n-1\}.
\end{equation}
Note that a smooth non-degenerate frame (coframe) field can be
defined on a manifold of a zero second Stiefel-Whitney class.
However this topological restriction is not exactly relevant
in physics because the solutions of physical field equations
can degenerate at a point or on a curve.
Moreover, these solutions produce the most important
physical models (particles, strings, etc.).\\
The coframe $\vt^a(x)$ represents, at a given point $x\in M$, a basis
of the linear space of 1-forms $\Omega^1$.
The set of all non-zero exterior products 
of basis 1-forms
\begin{equation}\label{A.3}
\vt^{a_1,\cdots,a_p}:=\vt^{a_1}\wedge\cdots\wedge \vt^{a_p}
\end{equation}
represents a basis of the linear space of $p$-forms $\Omega^p$.
Note the (anti)commutative rule for arbitrary forms $\a\in \Omega^p$
and $\b\in \Omega^q$
\begin{equation}\label{A.4}
\a\wedge\b=(-1)^{pq}\b\wedge\a.
\end{equation}
The dual set of vector fields
\begin{equation}\label{A.5}
\{e_a(x), \ a=0,\cdots,n-1\}
\end{equation}
forms a basis of the linear space of vector fields at a given point.\\
The duality of vectors and 1-forms can be expressed by {\it inter
product} operation for which we use the symbol $\hook$. Namely,
\begin{equation}\label{A.6}
e_a\hook\vt^b=\d^b_a.
\end{equation}
The action $X\hook w$ of  a vector $X$  on a form $w$ 
of arbitrary degree $p$ is defined by requiring: 
(i) linearity in $X$ and in $w$,  
(ii) modified Leibniz rule for the wedge product of
$\a\in \Omega^p$  and $\b \in \Omega^q$
\begin{equation}\label{A.7}
X\hook (\a\wedge\b)=(X\hook\a)\wedge\b+(-1)^p\a\wedge(X\hook\b).
\end{equation}
These properties together with (\ref{A.6}) guarantee the uniqueness of the
map $\hook:\Omega^p \to \Omega^{p-1}$.\\
The following relations involving the inner product operation ($p=deg(w)$)
are useful for actual calculations.
\brr\label{A.8}
X\hook(Y\hook w)&=&-Y\hook(X\hook w),\\
\label{A.9}
\vt^a\wedge (e_a\hook w)&=&pw,\\
\label{A.10}
e_a\hook(\vt^a\wedge w)&=&(n-p)w.
\err
We use also the forms $\vt_a:=\eta_{ab}\vt^b$ with subscript and the
corresponding vector fields $e^a:=\eta^{ab}e_b$ with superscript. Thus
\begin{equation}\label{A.11}
e_a\hook\vt_b=\eta_{ab}.
\end{equation}
The linear spaces $\Omega^p$ and $\Omega^{n-p}$ have the same dimensions
${n\choose k}={n\choose{n-p}}$. Thus they are isomorphic.
This isomorphism {\it Hodge dual map} is linear.
Thus it is enough
to define its action on basis forms:
\begin{equation}\label{A.12}
*(\vt^{a_1\cdots a_p})=\frac 1{(n-p)!}
\epsilon^{a_1\cdots a_pa_{p+1}\cdots a_n}\vt_{a_{p+1}\cdots a_n}.
\end{equation}
We use here the complete antisymmetric pseudo-tensor
$\epsilon^{a_1\cdots a_{n-1}}$ which is  normalized as $\epsilon^{01\cdots (n-1)}=1$.
The set of indices $\{a_1,\cdots,a_n\}$ is an even permutation of the
standard set $\{0,1,\cdots,(n-1)\}$.\\
Thus $*\vt^{0\cdots (n-1)}=1$ and $*1=-\vt^{0\cdots (n-1)}$.\\
The consequence of the definition (\ref{A.12}) is ($deg(\a)=deg(\b)$)
\begin{equation}\label{A.13}
\a\wedge *\b=\b\wedge\a.
\end{equation}
For the choice of the signature (\ref{A.1}) we obtain
\begin{equation}\label{A.14}
*^2w=(-1)^{p(n-p)+1}w.
\end{equation}
In the case $n=4$ the operator $*^2$ preserves the forms of  odd degree
and changes the sign  of the forms of  even degree.\\
The following equation is useful for actual calculations
\begin{equation}\label{A.15}
e_a\hook w=-*(\vt_a\wedge *w).
\end{equation}
To prove this linear relation it is enough to check it for
the basis forms.\\
The pseudo-orthonormality for the basis forms
$\vt^a$ yields the 
{\it metric tensor} $g$ on the manifold $M$ 
\begin{equation}\label{A.16}
g=\eta_{ab}\vt^a\otimes\vt^b.
\end{equation}   
The formulas (\ref{A.11}) and (\ref{A.15}) can be applied to derive
a useful form of a scalar product of two vectors $X$ and $Y$.
We write these vectors in the basis $e_a$ as $X=X^me_m$ and $Y=Y^me_m$.
Thus the scalar  product is
$$<X,Y>=X^mY^n<e_m,e_n>=X^mY^n\eta_{mn}.$$
Using (\ref{A.11}) we obtain
$$<X,Y>=X^mY^n(e_m\hook\vt_n)$$
Thus
\begin{equation}\label{A.17}  
<X,Y>=X\hook {}^\sharp Y=Y\hook {}^\sharp X,
\end{equation}
where ${}^\sharp X$ is the 1-form dual to the vector $X$ which obtained by 
a canonical map  from vectors to 1-forms
$$\sharp:X^me_m\to X^m\vt_m.$$
\section{Equivalence of (\ref{3.6}) and (\ref{4.24})}
Two forms of the field equation are linear in the coefficients
$\rho_i$.
Thus it is enough to prove the equivalence for separately for
every parameter.\\
Start with the $\rho_1$-terms by taking $\rho_2=\rho_3=0$ in both
equations.
The conjugated momentum (\ref{4.16a}) in this case
\begin{equation}
\F_a=\rho_1d\vt_a=\rho_1\C_a
\end{equation}
Insert this expression in the LHS of the equation (\ref{3.6})
\begin{eqnarray} \label{3-3}
&&
\rho_1\Big(
2d*d\vt_a+
e_a\hook(d\vt^b\wedge*d\vt_b)-
2(e_a\hook d\vt^b)\wedge*d\vt_b\Big)\nonumber\\
&&\qquad =2\Big[d*\F_a-\Big((e_a\hook \C^b)\wedge*\F_b
+\frac 12 e_a\hook(\C^b\wedge*\F_b)\Big)\Big]\nonumber\\
&&\qquad\qquad = 2(d*\F_a-\T_a)
\err
Consider the $\rho_2$-terms by taking $\rho_1=\rho_3=0$.
The congugated momentum (\ref{4.16a}) in this case
\begin{equation}
\F_a=\rho_2e_a\hook (d\vt^m\wedge \vt_m)
\end{equation}
Consequently by using (\ref{A.15})
\begin{equation}\label{l1}
*\F_a=\rho_2\Big(\vt_a\wedge* (d\vt^m\wedge \vt_m)\Big)
\end{equation}
and
\begin{equation}\label{l2}
d*\F_a=\rho_2\Big(d\vt_a\wedge* (d\vt^m\wedge \vt_m)
-\vt_a\wedge d* (d\vt^m\wedge \vt_m)\Big)
\end{equation}
Insert  the expressions (\ref{l1}) and (\ref{l2}) into the
$\rho_2$-term of the LHS of the equation (\ref{3.6}) to obtain
\brn
&&\!\!\!\!\!\!\!\!\!\!\!\!\!\!\!\!\rho_2\Big(
-2\vt_a \wedge d *(d\vt^b\wedge \vt_b)+
2d\vt_a \wedge * ( d\vt^b \wedge \vt_b)+\nonumber\\
&&\!\!
e_a\hook\Big(d\vt^c\wedge\vt_c\wedge*(d\vt^b\wedge\vt_b)\Big)-
2(e_a\hook d\vt^b)\wedge\vt_b\wedge*(d\vt^c\wedge\vt_c)
\Big)+\nonumber\\
&&\qquad =2\Big[d*\F_a-\Big((e_a\hook \C^b)\wedge*\F_b
+\frac 12 e_a\hook(\C^b\wedge*\F_b)\Big)\Big]\nonumber\\
&&\qquad \qquad =2(d*\F_a-\T_a)
\ern
As for the $\rho_3$-terms we take $\rho_1=\rho_2=0$.
The conjugated momentum (\ref{4.16a}) in this case
\begin{equation}
\F_a=\rho_3\Big(d\vt_a - \vt_a\wedge (e_m\hook d\vt^m)\Big)=
\rho_3e_m\hook (d\vt^m\wedge \vt_a)
\end{equation}
Consequently by using (\ref{A.15})
\begin{equation}\label{l3}
*\F_a=\rho_3\vt_m\wedge* (d\vt^m\wedge \vt_a)\Big)
\end{equation}
and
\begin{equation}\label{l4}
d*\F_a=\rho_3\Big(d\vt_m\wedge* (d\vt^m\wedge \vt_a)
-\vt_m\wedge d* (d\vt^m\wedge \vt_a)\Big)
\end{equation}
Insert  the expressions (\ref{l3}) and (\ref{l4}) into the
$\rho_2$-term of the LHS of the equation (\ref{3.6}) to obtain
\brn
&&\!\!\!\!\!\!\!\!\!\!\!\!\!\!\!\!\rho_3\Big(
-2\vt_b \wedge d*( \vt_a \wedge d \vt^b )+
2d\vt_b\wedge*(\vt_a\wedge d\vt^b)+\nonumber\\
&&\!\!
e_a\hook\Big(\vt_c\wedge d\vt^b\wedge*(d\vt^c\wedge\vt_b)\Big)-
2(e_a\hook d\vt^b)\wedge\vt_c\wedge*(d\vt^c\wedge\vt_b )
\Big)\\
&&\qquad =2\Big[d*\F_a-\Big((e_a\hook \C^b)\wedge*\F_b
+\frac 12 e_a\hook(\C^b\wedge*\F_b)\Big)\Big]\nonumber\\
&&\qquad = 2(d*\F_a-\T_a)
\ern
Consequently the equivalence of two forms of the field equation is
proven for all values of the parameters.


\end{document}